\documentclass[english,12pt]{article}

\usepackage[latin1]{inputenc}
\usepackage[T1]{fontenc}
\usepackage{amsmath}
\usepackage{amsfonts}
\usepackage{graphicx}
\usepackage{subfigure}
\usepackage{a4wide}
\usepackage{amssymb}
\usepackage{fancyhdr}
\usepackage{mathrsfs}
\usepackage[toc,page]{appendix}

\begin{document}


\title{Black holes, wormholes and compact stars in Spherically Symmetric Space-times}

\author{ 
Lorenzo Sebastiani\footnote{E-mail address: lorenzo.sebastiani@unitn.it
},\,\,\,
Luciano Vanzo\footnote{E-mail address: luciano.vanzo@unitn.it},\,\,\,
Sergio Zerbini\footnote{E-mail address: sergio.zerbini@unitn.it}\\
\\
\begin{small}
Dipartimento di Fisica, Universit\`a di Trento,Via Sommarive 14, 38123 Povo (TN), Italy
\end{small}\\
\begin{small}
TIFPA - INFN,  Via Sommarive 14, 38123 Povo (TN), Italy
\end{small}
}



\date{}

\maketitle

\abstract{We present an unified approach for the study of idealized gravitational compact objects like wormholes and horizonless stars, here simulated by the presence of boundary conditions at a deeply inner wall. At classical level, namely neglecting quantum effects,  the presence of the 
wall leads to characteristic echoes following the usual ringdown phase, such that it can discriminate black holes from other horizonless, and probably exotic, compact objects. 
With regard to this issue, an analytical though approximated expression for the complex frequencies of the quasinormal echoes is found and discussed. At quantum level, we show that static wormholes do not radiate.}

\section{Introduction}

Recently, the first detection of gravitational waves from binary systems of black holes (BHs) and the ``multimessenger'' 
signals obtained from the first observation of the collision of two
relativistic neutron stars have led to important progresses in the understanding of the most massive and compact objects of the Universe. 
In these events, strong  gravity effects play a fundamental role with respect to the other interactions known in nature. The predictions of General Relativity (GR) 
have been substantially confirmed: thanks to the data coming from LIGO~\cite{LIGO1,LIGO2, LIGO3, LIGO4}, the existence of the gravitational waves has been confirmed.
Within the experimental error, the gravitational waves signals have been related to the most  compact macroscopic objects in nature, namely BHs, 
but the possible existence of  less extreme compact objects could  not be rule out. In fact, we recall that BHs lead to ringdown waveform completely 
determined by quasinormal modes (QNMs), which depend only on the mass and angular momentum of BH. Thus, the possible presence of additional ``echoes'' 
in the ringdown waveform might be the evidence of the absence of event horizon. As a consequence,   exotic horizonless compact objects (ECOs) (gravastars~\cite{grav, grav2}, bosonstars~\cite{bs}, or other exotic compact objects ~\cite{Mark, ECO1, ECO2, Conk, Ma, ECO3, ECO4}), could not be excluded as alternative to BHs. True, such objects ought to violate the Buchdahl stability limit, since $R<2M<9M/4$ in order that echoes be present, and thus be formed from exotic forms of matter. Wormholes are a possibility here, although they  cannot be considered as compact objects in the usual sense. Therefore we distinguish between ECOs and wormholes, also on the ground that many wormhole solutions are exactly known in GR without making appeal to other theories.

Furthermore, since we do not know the metric associated with  a gravitational collapse of two BHs, also in this case echoes could be present. 
Finally, we recall that several speculations have been proposed (see Ref.~\cite{rev000} and references therein for a recent review and Ref.~\cite{As1}
 for the search of echoes in LIGO data) 
concerning the behaviour of the space-time near to the 
BH horizon and its connection to possible quantum gravity effects or alternative theories of gravity~\cite{Kon1}.

On general ground, the gravitational ringdown waveform related to an  ECO contains QNMs similar to BHs, 
but with additional  sequences of distinct pulses compatible with echoes. This peculiar behaviour takes place also if one considers the geometry of a wormhole
(WH)~\cite{Morris,visser}, where 
the spatial region is bounded, but the event horizon is not present.

In this paper, we will try to present an analysis, investigating at the same time  some features of static spherically symmetric BHs, ECOs and WHs. 
In most cases concerning BHs and WHs we are going to study, we will try to show that even if the metrices differ between them by small parametrs, their nature can be very different,

In particular, we will derive, making use of WKB techniques and in an analytical way,  the spectrum of frequencies forming (part of) the echoes spectrum in the post ringdown phase, when 
the event horizon is not present (ECOs and WHs) and the matter replacing it ceased to oscillate. One hopes that future analysis of observational data could discriminate between BHs and other objects. \\
In order to be clear on what we are talking about, we must be careful about terminology. In the literature, quasi-normal modes are special black hole's perturbations characterised by complex frequencies such that the corresponding wave form decays in time. The post merger signal (including echoes) for realistic sources has undoubtedly a more complicated pattern, resulting in part from a time dependent source and also in part from quasi-normal ringing. The distinction has been clearly emphasised recently by Price and Khanna\cite{Price:2017cjr}, where a problem similar to ours has been considered. In this work we do not consider the former part, but only the spectrum of the possible frequencies after the compact object has reached a stationary state (if it does), and we agree to call these frequencies quasi-normal modes by extension of the black hole case, although only ``half'' of the boundary conditions of the black hole case are strictly employed (outgoing waves are only present in the exterior region).  \\
Whether these frequencies really appear as components in the final signal is a non trivial question, due to the lack of completeness of the modes for a non self-adjoint problem. In fact, in Nollert\cite{Nollert:1996rf} it is shown that a small modification of the curvature potential has great effects on the QNM spectrum, but a little one on the evolution of the Cauchy data. 
However, even if completeness is lost, for potentials with compact support there is a theorem of ``asymptotic completeness'' due to Lax and Phillips in their treatment of scattering theory\cite{Lax1989}, according to which the asymptotic form of the signal as $t\to\infty$ is fully determined by all QNM resonance frequencies and the anti-bound states (those purely imaginary frequencies with $\rm{Im}\,\omega<0$); that is for a perturbation $\Phi(t,x)$, solution of the Regge-Wheeler (or similar) equations

\begin{equation}
\Phi(t,x)\approx\sum_{j}C_{j}e^{-i\omega_{j}t}f_{j}(x,\omega_{j}), \qquad t\to+\infty
\end{equation}
where $f_{j}(x,\omega_{j})$ are the Jost functions of the scattering problem in the curvature potential of the black hole or the compact object. It is reasonable that the theorem holds good in the non compact case, if the potential is rapidly decreasing at $r\to\infty$, and results in this direction are known after the important work of Bachelot and Bachelot\cite{Bachelot:1993dp}. Therefore it is clear that the resonant frequencies computed in Section [IV] do appear in the asymptotic signal, the only omission of the calculations being the anti-bound states (which do not oscillate at all but only dump) and the scattering part belonging to the part of the spectrum  which is ``over the barrier'' of the curvature potential.   There are no other sources of signals for this particular, idealised, stationary, compact object replacing the black hole, so after the source has ``settled down'' this is all we have according to the asymptotic theorem.

At the fundamental level, we also present a result concerning the Hawking radiation. It is well know
that BHs are not black at quantum level, but the Hawking thermal radiation is present. Of course ECOs do not lead to Hawking radiation, but we will show that also static WHs 
do not radiate in the Hawking sense. Only in the dynamical case, particle creation can be present also in the WHs, but in the dynamical case, the distinction 
between BHs and WHs is very subtle. 

The paper is organized in the following way. In Section {\bf 2} we will revisit the covariant formalism which permits the description of a generic Spherically Symmetric 
Space-time. In the static case,
this formalism  gives an unique way to define BHs and  horizonless objects as ECOs and WHs. 
In Section {\bf 3} we will investigate fluid models for BHs, ECOs and WHs in the framework of GR. In particular, we will propose a reconstruction of 
regular BH models where the instabilities related to the (popular) choice of  negative radial pressure may be removed. Section {\bf 4} 
is devoted to the study of QNMs and echoes, and here the analytical formula is presented.
In Section {\bf 5} we will add few considerations to the Hawking radiation of both BHs and WHs. 
Conclusions and final remarks are given in Section {\bf 6}. Several Appendices are also added, where auxiliary material is presented.

In this paper, we set the Newton's constant as $G_N=1$.

\section{Spherically Symmetric Space-times}

To begin with, for the sake of completeness, 
we recall the invariant formalism we shall make use of (see for example Refs.~\cite{sean09,noi,vanzo}), 
valid for a generic four dimensional Spherical Symmetric (SS) space-time,
\begin{equation}
ds^2=\gamma_{ab}dx^a dx^b+ r(x^a)^2 dS^2\,,
\end{equation}
where $dS^2$ represents the metric of a two-dimensional sphere, but the formalism can be easily generalized to deal with a two-dimensional maximally symmetric space, 
$\gamma_{ab}$ is 
the metric tensor of the two dimensional space-time (the normal metric),  with coordinates $x^a\,, a=0,1$, and $r\equiv r(x^a)$ is the areal radius and is a 
function of the coordinates 
of the normal space-time. 
Furthermore, on the normal space-time, it is possible to introduce the scalar quantity,
\begin{equation}
 \chi(x^a)=\gamma^{a b}\partial_a r\partial_b  r\,,\label{chi0}
\end{equation}
which defines a (dynamical) trapping horizon in the following way,
\begin{equation}
\chi(x_H)=0\,,\quad \partial_a\chi(x_H)>0\,,\quad a=0,1\,. \label{Hcond}
\end{equation}
The Hayward surface gravity~\cite{sean09}, a scalar quantity related to the existence of the trapping horizon, is  
\begin{equation}
\kappa_H= \frac{1}{2\sqrt{-\gamma}} \partial_a\left(\sqrt{-\gamma}\gamma^{ab} \partial_b r\right)_H \,.\label{kHay}
\end{equation}
Finally, the last important quantity which 
can be introduced in a generic SS space-time is the Kodama vector \cite{kodama}. In our notation it is the vector on the normal space given by
\begin{equation}
K^a=\frac{\varepsilon^{ab}}{\sqrt{-\gamma}}\partial_ar \,.\label{K}
\end{equation}
According with the definition, it follows that the Kodama vector is covariant conserved in a generic SS space-time.

\subsection{The static case}

The Static spherically Symmetric (SSS) case is well understood and
we will see that, in the presence of horizons, the Kodama vector allows an invariant way to distinguish between black holes and wormholes, apart
from the case of horizonless compact objects. Here we will follow the unified and general approach proposed by Sean Hayward \cite{seanWH}.

It is convenient to start with the (usual) metric with coordinates $(t, \rho)$ in the normal space, namely
\begin{equation}
ds^2=-A(\rho)dt^2+\frac{d\rho^2}{B(\rho)}+ r^2(\rho) dS^2\,,
\label{sgen0}
\end{equation}
where $A\equiv A(\rho)$ and $B\equiv B(\rho)$ are functions of the radial coordinate $\rho$ only.
An important example is  the diagonal Schwarzschild gauge with $r=\rho$,
\begin{equation}
ds^2=-A(r)dt^2+\frac{dr^2}{B(r)}+ r^2 dS^2\,.
\label{sg00}
\end{equation}
The associated static isotropic form reads
\begin{equation}
ds^2=-A(\rho)dt^2+\frac{r^2(\rho) d\rho^2}{\rho^2}+ r^2(\rho) dS^2\,,
\label{sgiso0}
\end{equation}
where
\begin{equation}
\frac{ d\rho}{\rho}=\frac{dr}{r\sqrt{B}}\,.
\label{sgiso}
\end{equation}
Another example is the Regge-Wheeler gauge
\begin{equation}
ds^2=A(\rho^*)\left(-dt^2+d\rho^{*2}\right)+ r^2(\rho^*) dS^2\,,
\label{rw}
\end{equation}
with the tortoise given by
\begin{equation}
d\rho^*=\frac{ d\rho}{\sqrt{AB}}\,.
\label{sgiso2}
\end{equation}
We also recall the Painlev\`e gauge,
\begin{equation}
ds^2=-A(\rho)dv^2+2\sqrt{\frac{A(\rho)}{B(\rho)}}dvd\rho+ r^2(\rho) dS^2\,,
\label{pg0}
\end{equation}
where the temporal coordinate follows from the identification  $dv=dt+d\rho^*$. 

Finally, the proper distance (or Ellis) gauge reads
\begin{equation}
ds^2=-A(\sigma)dt^2+d\sigma^2+ r^2(\sigma) dS^2\,,
\end{equation}
with
\begin{equation}
d\sigma=\frac{d\rho}{\sqrt{B}}\,.
\label{pd}
\end{equation}
Within the general gauge (\ref{sgen0}), the invariant quantity $\chi\equiv \chi(x^a)$ in (\ref{chi0}) reads
\begin{equation}
\chi=B(\rho)\left(\frac{dr}{d\rho}\right)^2\,.
\label{chi}
\end{equation}
The Kodama vector (\ref{K}) is given by 
\begin{equation}
K=\left( \sqrt{\frac{B}{A}}\frac{d r}{d\rho},0,0,0 \right)\,.
\label{k1}
\end{equation}
Furthermore, one may introduce the so called Kodama energy $\omega$ associated with a test particle with four-momentum  $p_\mu=\partial_\mu I$, 
where $I$ is the relativistic action. Thus, making use of the Kodama vector (\ref{K}),
\begin{equation}
\omega=-K^\mu p_\mu= \sqrt{\frac{B}{A}}\frac{d r}{d\rho} E\,,
\label{k20}
\end{equation}
where   $E=-\partial_t I$ is the test particle Killing energy. Alternatively, if $T_{\mu \nu}$ is a matter stress-energy tensor, we may consider the four-vector
\begin{equation}
Q_\mu=T_{\mu \nu}K^\nu \,,                                         
\label{k3}
\end{equation}
and in order to deal with  a scalar quantity, we may introduce 
$\Sigma=K^{\mu}Q_{\mu}=T_{\mu \nu}K^\nu K^{\mu}$. In the static case, one has
\begin{equation}
\Sigma=\frac{B}{A}\left(\frac{d r}{d\rho}\right)^2 T_{00}\,.
\label{k220}
\end{equation}
By definition, both BHs and WHs possess a trapping (event) horizon, given by $ \chi_H= 0$, and this has to be a simple zero. On the other side, if $\chi$ is
never vanishing, one obtains
a ECO or a static solution with a naked singularity.

In the above general static gauge, the trapping horizon is given by,
\begin{equation}
 \left(B(r)\left(\frac{d r}{d\rho}\right)^2\right)_H=0 \,,
\label{sth}
\end{equation}
which simplifies when $\rho=r$ as,
\begin{equation}
B(r_H)=0 \,.
\label{sths}
\end{equation}
Let us assume that there exists a trapping horizon. Then, we could define a {\it static black hole} as a SSS solution such that  
$\omega_H$ (or $\Sigma_H$) is {\it not vanishing}. 
By looking at (\ref{k20}), it is clear that this is possible if
one {\it also} has $A(r_H)=0$, with $A'(r_H) >0$ (we will denote with the prime index the derivative with respect to the radial coordinate $r$). 
In this case, $A(r)$ has to be proportional to $B(r)\left(\frac{d r}{d\rho}\right)^2 $. 
In the gauge where $r=\rho$, for a BH, we find $B(r_H)=0$, 
with $B'_H >0$, and, due to the fact that  $A(r)$ is proportional to $B(r)$, $A(r_H)=0$, with $A'(r_H) >0$. 
The Hayward surface gravity (\ref{kHay}) associated with a trapping horizon is defined  as
\begin{equation}
\kappa_H=\frac{B'_H}{2}\,, 
\end{equation}
with the Killing surface gravity given by
\begin{equation}
\kappa_H^{k}=\frac{\sqrt{B'_H A'_H}}{2}\,.
\label{k2}
\end{equation}
For most of known BHs, $A(r)=B(r)$, and the two surface gravities coincide.

On the other hand,  if  $\omega_H$ (or $\Sigma_H$) is {\it vanishing}, we  are in the presence of a {\it static wormhole}.  
It means that $A(r_H)\neq 0$ on the horizon. In this case, 
according to the Hayward classification one has to deal with a  {\it double trapping horizon}. 
The Hayward surface gravity reads
\begin{equation}
\kappa_H=\frac{B'_H}{4} \left(\frac{d r}{d\rho}\right)_H \,,
\label{k22}
\end{equation}
when $r \neq \rho$, and 
\begin{equation}
\kappa_H=\frac{B'_H} {4}\,, 
\label{k222}
\end{equation}
when $r=\rho$.
The positivity of the Hayward surface gravity means that there exists a minimum value for $r$, which gives rise to the existence of a {\it throat or mouth of a 
traversable wormhole}. 

\subsection{Some examples}

Let us start with the exponential gravity metric (see the recent paper \cite{Visser18} where original references can be found),
\begin{equation}
ds^2=-e^{-\frac{2m}{\rho}} dt^2+\frac{d\rho^2}{e^{-\frac{2m}{\rho}}}+ \rho^2 e^{\frac{2m}{\rho}} dS^2\,,
\label{sgenturco}
\end{equation}
where $m$ is a mass constant.
Here the areal radius is $r=\rho e^{\frac{m}{\rho}}$. Thus, making use of (\ref{sth}), the trapping horizon is located at $ \rho_H=m$ and the associated surface gravity reads
\begin{equation}
\kappa_H=\frac{e^{-1}} {2m}>0\,.
\label{k13}
\end{equation}
It means that one is dealing with  a double trapping horizon since also $A(r_H)>0$, and $\omega_H=0$ and we have a traversable static WH, as first observed  in \cite{Visser18}. 
However, this metric is not phenomenological viable.

Another interesting metric is the one proposed by Damour and Solodhukin \cite{DS} given by,
\begin{equation}
ds^2=-\left(1-\frac{2m}{r}+\lambda^2\right)dt^2+\left(1-\frac{2m}{r}\right)^{-1}dr^2+r^2 dS^2\,,
\label{DS}
\end{equation}
where $m$ is still a mass constant and
$\lambda$ is an arbitrary (but small) real constant. Here $\rho=r$, the trapping horizon is $r_H=2m$, and one has to deal with a WH, since $A(r_H)=\lambda^2>0 $ 
and the Kodama energy vanishes ($\omega_H=0$).

A variant we shall discuss, and which is asymptotically Minkoskian may be defined  by
\begin{equation}
ds^2=-\left(1-\frac{2m(1-b^2)}{r}\right)dt^2+\left(1-\frac{2m}{r}\right)^{-1}dr^2+r^2 dS^2\,,
\label{DSZ}
\end{equation}
where $b^2$ is a dimensionless very  small parameter. Here, again, $r_H=2m$,  but $A(r_H)=b^2 >0$ and arbitrarily small.

\section{Fluid Models in General Relativity}

We are interested in solutions of Einstein equation in presence of perfect relativistic anisotropic fluids. This issue has been largely studied in the literature. 
For the isotropic perfect fluid see Ref.~\cite{Lake}. In the following, we only make some remarks on BHs,  WHs and ECOs.

Starting from a SSS metric in the diagonal Schwarzschild gauge,
\begin{equation}
ds^2=-A(r)dt^2+\frac{dr^2}{B(r)}+ r^2 dS^2\,,
\label{sg0}
\end{equation}
it may be convenient to pass to Painlev\`e gauge (\ref{pg0}), introducing the advanced time $dv=dt+ \frac{dr}{\sqrt{AB}}$.
Thus,
\begin{equation}
ds^2=-A(r)dv^2+2\sqrt{\frac{A(r)}{B(r)}}dvdr+ r^2 dS^2\,.
\label{pg}
\end{equation}
The Einstein equations are
\begin{equation}
G^{\mu \nu}=8\pi T^{\mu \nu}\,,
\label{ee0}
\end{equation}
with
\begin{equation}
 T^{\mu \nu}=(\rho+p_T)u^\mu u^\nu+p_Tg_{\mu \nu}+(p_r-p_T)C^\mu C^\nu\,,
\label{ee}
\end{equation}
where $u^\mu u_\mu=-1$ is a time-like vector and $C^\mu C_\mu=1$ the anisotropy space-like vector. 
It is clear that $\rho$ is the energy density, $p_r$ the radial pressure and $p_T$ the transversal pressure of the fluid.
One derives the following equations,
\begin{equation}
r B'+B-1=-8\pi r^2 \rho\,,\label{b}
\end{equation}
\begin{equation}
\frac{A'}{A}-\frac{B'}{B}=\frac{8\pi r (\rho+p_r)  }{B}\,,\label{ab}
\end{equation}
\begin{equation}
  p_r'+\frac{\rho+p_r}{2}\frac{A'}{A}=
  \frac{2 (p_T-p_r  )}{r}   \,.\label{abc}
\end{equation}
Thus, given the energy density $\rho$,  $B$ can be computed from (\ref{b}). 
Once an equation of state is chosen, namely $p_r=f(\rho)$, the metric function $A$ can be also evaluated from (\ref{ab}). 
Finally, the third equation, Tomlman-Oppenheimer-Volkov equation, gives the form of $p_T$.

As it is well known in vacuum, i.e. $\rho=p_r=p_T=0$, the only solution is the Schwarzschild one, namely
\begin{equation}
A(r)=B(r)=1-\frac{2M}{r}\,,\label{sw}
\end{equation}
where $M$ is an integration constant, identified with the black hole mass. 
In this celebrated case, the trapping horizon coincides with the so called event horizon and it is given by the simple zero of $\chi=B(r)=0$, 
namely $r_H=2M$. 
Here, the Hayward surface gravity coincides with the usual Killing surface gravity, $\kappa_H=B'_H/2=(4M)^{-1}$. 
Furthermore, the Kodama vector also coincides with the Killing vector $(1, 0,0,0)$.

For illustrative purposes we mention a (rather unphysical) static  WH solution we have already discussed in (\ref{DSZ}) is,
\begin{equation}
B(r)=1-\frac{2m}{r}\label{sw1}\,,
\end{equation}
\begin{equation}
A(r)=1-\frac{2m_1}{r}\label{sw2}\,,\quad m_1=m(1-b^2)\,,
\end{equation}
In this case, one has $\rho=0$, while the radial pressure reads
\begin{equation}
4\pi r^2 p_r=-\frac{mb^2}{r-2m_1}<0\label{swd3}\,.
\end{equation}
Here, $r_H=2m$ and  $A(r_H)=b^2$, with again $\omega_H=0$. 
The form of $p_T$ can be computed and it is easy to see that the Null energy Condition (NEC ) is violated, which is consistent with the WH form of the metric. 
Several other examples can be constructed. In the Appendix A, it is reviewed another way to deal with possible 
existence of static wormholes, namely  scalar tensor models.

Within the framework of GR and by following a similar strategy, one may also discuss the so called regular  BH solutions \cite{Bardeen, seanBH}, and for a review see 
Refs.~\cite{ans,bronnikov, CZ, CCZ}, and their horizonless counterpart. 

In this respect, a popular choice is the equation of state $p_r=-\rho$. This gives $A=B$, and $B$ is determined by solving (\ref{b}).

For example, making the choice
\begin{equation}
B(r)=A(r)=1-\frac{2mr^2}{r^3+2l^2m}\label{h1}\,,
\end{equation}
$m\,,l$ being constants,
one has  the well known Hayward regular black hole~\cite{seanBH}, which has a de Sitter core for small $r$ and reduces to
the Scharzschild case for large $r$. We recall that in order to deal with a BH, the mass parameter $m$ has to satisfy the inequality $m>m_c=3^{3/2}l/4$. For $m< m_c$, one 
has a horizonless ECO. Other similar solutions are the so called gravastars~\cite{grav}. 

Thus, the regularity stems from the presence of the parameter $l$: once it is not vanishing, the central singularity is cured because there exists a de Sitter core. 
However, here the problems are two: the first one is that the radial speed of sound
$ v^2=\frac{d p_r}{d \rho}=-1 <0$, a clear signal of instability. The second one, is associated with the presence of a Cauchy horizon, with relative instability 
related to mass inflation
(see the recent review \cite{mi}).
In order to avoid such kind of instabilities, one may try to investigate  another equation of state. 
The simplest one is represented by $p_r=0$. With this choice,  one finds that the Einstein equations are solved by
choosing $A$, and with $B$ given by
\begin{equation}
B(r)=\frac{A(r)}{rA'(r)+A(r)}\,\label{hd1}\,.
\end{equation}
The related energy density reads
\begin{equation}
8\pi \rho=\frac{A(r)}{r}\frac{2A'(r)+rA''(r)}{(rA'(r)+A(r))^2}\label{hdr}\,.
\end{equation}
First, it should be noted that no static WH solutions exist within this choice $p_r=0$, since if $B(r_H)=0$ 
on the horizon, it follows that $A(r_H)=0$, provided that $A'_H$ is not vanishing. For example, if we make the choice $A(r)=1-\frac{2mr^2}{r^3+2ml^2} $, 
one gets an Hayward ``dirty'' ($A\neq B$) regular BH, and if the mass parameter is sufficiently small, the related ECO. 
Note that $\rho$ is positive only when $A(r)>0$. Thus, for a horizonless ECO for which $A(r)$ is always positive,  $\rho >0$. Furthermore, $p_T$ can also be computed, and reads
\begin{equation}
p_T=\frac{1}{32}\left(\frac{(2A'(r)+rA''(r))A'(r)}{(rA'(r)+A(r))^2} \right)\label{hd99}\,.
\end{equation}

We conclude with a remark concerning BHs with $p_r\neq-\rho$. Since on the horizon only $A(r_H)=0$, $p_T$ is finite here.

As a consequence, with equation of state $p_r=0$, one has a class of dirty black holes, 
generated by a fluid  with density and radial pressure vanishing on the horizon. Since the energy density is positive for $r> r_H$, it follows that in the interior where $r<r_H$ 
the energy density is negative, but {\it finite} at $r=0$. In fact, for small $r$, we get $A(r)=1-ar^2$, namely a de Sitter core, the same behavior for $B(r)$. 
Now, making use of equation (\ref{hdr}), it follows that $\rho$ is finite at $r=0$. However, the problematic Cauchy horizon is still  present.


\section{Quasinormal modes and echoes}

\textit{The Perturbation Equation} - In this Section, we shall discuss the QNMs (see for example Ref.~\cite{kokko}) related to the static solutions we have discussed, using  the general background metric 
\begin{equation}
ds^2=-A(\rho)dt^2+\frac{d\rho^2}{B(\rho)}+ r^2(\rho) dS^2\,.
\label{sgen}
\end{equation}
We consider a massless scalar perturbation $\Phi(t,\rho, \tilde\Omega)$ on the given background metric, obeying to the equation,
\begin{equation}
\frac{1}{\sqrt{-g}} \partial_{\mu}\left(\sqrt{-g} g^{\mu \nu}\partial_\nu \right) \Phi(t,\rho, \tilde\Omega)=0 \,.
\label{sgenp0}
\end{equation}
Here, $\tilde\Omega=(\theta,\phi)$ represents the angular coordinates.
Making the standard separation of variables,
\begin{equation}
\Phi(t,\rho, \tilde\Omega)=   \frac{1}{\rho} \phi(\rho)Y_{lm}(\tilde\Omega) e^{-i\omega t} \,,
\label{sgenp}
\end{equation}
where $Y_{lm}(\tilde\Omega)$ are the spherical harmonics, $\phi\equiv\phi(\rho)$ depends on the radial coordinate only and $\omega$ is the (generically complex) frequency  of the perturbation. With this choice, we are interested in solutions with $ \mbox{Im}\, \omega >0$. Thus, one arrives at
\begin{equation}
-\frac{d^2 \phi}{d^2 \rho^*} +V(\rho)\phi=\omega^2 \phi \,,
\label{rwe}
\end{equation}
where the tortoise coordinate $\rho^*$ is defined by (\ref{sgiso2}) such that $\rho=\rho(\rho^*)$
and for scalar perturbation the effective potential reads
\begin{equation}
V(\rho)=\frac{A(\rho)(l(l+1))}{r^2(\rho)}+\frac{1}{r(\rho)}\frac{d^2 r}{d \rho^{*2}} \,,
\label{effp}
\end{equation}
$l$ being  the orbital angular momentum quantum number. For a massless spin $s$, one has (see for example  \cite{kono11}),
\begin{equation}
V(\rho)=\frac{A(\rho)(l(l+1))}{r^2(\rho)}+\frac{1-s^2}{r(\rho)}\frac{d^2 r}{d \rho^{*2}} \,.
\label{effps}
\end{equation}
Thus, for electromagnetic perturbation with $s=1$, only the first term is present. 
We should note that for this last case with $s=1$ or for large values of $l$,
the maximum of the potential corresponds to the photosphere (see Appendix B), where the condition
\begin{equation}
r\frac{d A(\rho)}{d \rho}-2A(r)\frac{d r}{d\rho}=0\,, 
\end{equation}
has met. In any case, the effective potential associated to BHs, WHs and ECOs possesses a local maximum for positive values of the tortoise.  
When $\rho^*$ is negative, the potential tends to vanish in the case of BHs, while reaches a minimum for an ECO and on the throat of a WH, defined as the location of the minimum area space-like two-sphere in the manifold. Finally, for larges values of $\rho^*$ the effective potential goes to zero (see also Appendix C).

As a check, in the case of Schwarzschild BH with $\rho=r$, $A=B=1-2M/r$, $M$ mass constant, we have
\begin{equation}
- \frac{d^2 \phi}{d^2 r*} + \left(1-\frac{2M}{r}\right)\left(\frac{l(l+1)}{r^2}+(1-s^2)\frac{2M}{r^3} \right) \phi =\omega^2 \phi\,,
\label{rwesc}
\end{equation}
which coincides with the well known equation.

Coming back to  the perturbation equation (\ref{rwe}), 
we recall that it is an example of symmetric Sturm-Liouville problem, namely the related second order differential operator is symmetric. 
All the difference between the black hole and the various ECOs lies in the numerical range of $\rho_{*}$. It is the real line for BHs,
it is a segment $a\leq \rho_{*}$ for ECOs. But in this case we will assume that $a$ is well inside the peak of the potential barrier surrounding the collapsed object,
hence well inside the light sphere. And furthermore the ECO will be effectively described by a boundary condition at $\rho_{*}=a$, which summarizes our ignorance on 
the nature of the object. 
A discussion about QNMs and echoes in terms of boundary conditions can be found in Ref.~\cite{cardoso},
where the QNMs of a double-barrier wormhole have been computed and compared to those of a Schwarzschild BH (see also Ref.~\cite{Bueno}).

First, let us formulate the problem in general terms. In order to deal with BHs,  WHs, and ECOs it is notoriously convenient to make use of the tortoise coordinate related to the radial coordinate $r$. Thus, given the static metric

\begin{equation}
ds^2=-A(r)dt^2+\frac{dr^2}{B(r)}+ r^2 dS^2\,.
\label{sg}
\end{equation}
the tortoise coordinate, denoted here by $x$, is
\begin{equation}
x=\int \frac{dr}{\sqrt{AB}}\,.
\label{xsg}
\end{equation}
In this Section, the prime index will be the derivative with respect to $x$.
As a consequence, for BH, the range $(r_H, \infty)$ is mapped in the whole real line  $(-\infty, \infty)$. For WH, the range  $(r_H, \infty)$ is mapped in   
$(x_H, \infty)$, with $x_H<<0$ in mass units. For the WH case, the use of a boundary condition at the throat correspond to working in the manifold $M/Z_{2}$, where $M$ is 
the wormhole and $Z_{2}$ is the isometric reflection map of $M$ (see Appendix C)\footnote{There are other ways to extend the asymptotically flat region beyond the 
throat, which we do not consider here.} 

For example, in a variant of Damour-Solodukhin  WH in (\ref{sw1})--(\ref{sw2}), with $m_1=m(1-b^2)$, we get
\begin{equation}
x=\sqrt{(r-2m)(r-2m_1)}+2(m+m_1)\ln\left( \frac{\sqrt{r-2m_1}+\sqrt{r-2m}}{{2\sqrt{2m}}}  \right)\,.
\label{xds}
\end{equation}
As a result, on the horizon $r=2m$, one has $x_H=2m\left(1-\frac{b^2}{2}\right)\ln\left[\frac{b^2}{4}\right]$. Thus, if $b^2<<1$, $x_H$ assumes very large negative values in units of mass.
Not surprisingly, when $b^2=0$ one obtains the well known BH result
\begin{equation}
x=r-2m+2m\ln\left( \frac{r-2m}{2m}  \right)\,,
\label{sxds}
\end{equation}
and the horizon is mapped in $x_H\rightarrow -\infty$.

In the case of ECOs, we shall  investigate only  a restricted class, namely compact objects whose metric is a ``quasi regular black hole'' in the Sakharov sense, namely having a de Sitter core for small $r$, but $A(r)=B(r)$ is never vanishing, and {\it there is no event horizon, nor a related Cauchy horizon}.
It means that a local minimum is present when
$0<A(r)<<1$. It follows that the range   $(0, \infty)$ is mapped in   $(x_c, \infty)$, with $x_c<<0$. 
One example is the Hayward metric, 
\begin{equation}
A(r)=B(r)= \frac{r^3+2l^2m-2mr^2}{r^3+2ml^2}=\frac{(r+a)((r-b)^2+c^2)}{r^3+2ml^2}\,,
\label{hco}
\end{equation}
where we have parametrized the numerator with the negative real root $r_1=-a$\,, $a>0$, and the two complex roots $r_{\pm}=b \pm i c$, with $b\,,c$ real numbers. 
In this way, $A(r)>0$ for $r>0$. We recall that this happens only if $m<3^{3/2}l/4$. This condition is model dependent. For example, in the Balart-Vagenas-CRKB 
example~\cite{Bal,Co},
\begin{equation}
A(r)=B(r)= \frac{(r +l_v)^3-2mr^2}{(r+l_v)^3}\,,
\label{hco1}
\end{equation}
where $l_v=\frac{2m}{L}$, with $1/L$ related to quantum corrections, one has  $A(r)>0$ for $r>0$ as
soon as $m>\frac{2L}{27}$, recalling that we are using $G_N=c=1$ units.

In all these cases, it is possible to compute the tortoise $ x $. The result is
\begin{equation}
x=r+\frac{1}{c} \tan^{-1}\left[\frac{r-b}{c}\right]G(r)+F(r)\,,
\end{equation}
where $G(r)$ and $F(r)$ are known functions. When $r \rightarrow \infty$, then  $x \rightarrow \infty$. Furthermore,
\begin{equation}
x(0)=-\frac{1}{c} \tan^{-1}\left[\frac{b}{c}\right]G(0)+F(0) \,.
\end{equation}
Thus, if $0<c<<1 $, $x(0)\simeq-\frac{1}{c} <<0$, and the range is similar to the one of WH case.

\textit{Echoes and WKB} - Now the usual boundary conditions (B.C.) associated with the QNMs for the BH 
are the Sommerfeld B.C., namely {\it outgoing plane waves at spatial infinity $\rho^* \rightarrow +\infty$} and {\it ingoing plane waves at horizon}, 
since in the BH case, the horizon is reached when  
$\rho^* \rightarrow -\infty$.
It is possible to show that such B.C. make  the resolvent of the perturbation operator {\it singular}. 
As a result, the  singular values of $\omega$ are {\it complex numbers}, namely the chosen boundary conditions are dissipative. Several analytic 
approximation techniques are at disposal.

We proceed by considering at first the cases of WHs and ECOs and at second by taking the BH limit of the results.
Coming back to the equation for the perturbation, one has to deal with 
\begin{equation}
L\, \phi=\frac{d^2 \phi}{d x^2}+(\omega^2-V(x))\phi=0 \,.
\label{11r}
\end{equation}
Thus, we have formally a Sturm-Liouville problem, and it is convenient to work in the Hilbert space $L_2(x_c, \infty)$, with $x_c=x_H$ for WH and $x_c=x(0)$ for a ECO, both negative and very large. It can be considered in some sense as a stretched horizon, once popular in the black hole community after 't Hooft made use of it for quantum black holes. One may wish to avoid an interior boundary in the case of a static wormhole, and work instead with the two symmetrical barriers equidistant from the throat, but the basic reasons for echoes is just the same, namely trapped modes which eventually escape the throat. In all cases,  the perturbation operator $L$ is symmetric.\\
 Furthermore, we know that the effective potential $V(x)$ is vanishing at $|x|$ very large, at least as $x^{-4}$, and the first derivative even faster, as $x^{-5}$ (we want the object to look as far as possible as the exterior region of a non rotating black hole). Thus the potential flattens and vanishes on the external side and admits a single local maximum at some $x_M>0$, of order $\alpha M^{-2}$ in Planck units. This maximum raises as the angular momentum raises while preserving the shape of the potential, in particular its width.  On the inner side with respect to the potential wall, it vanishes exponentially fast for all metrics described so far, with derivatives vanishing accordingly. Thus, for $x_c <x\ll0$ (and $x\gg x_{M}$) we are in a valid WKB regime, and one may try the approximate solutions
\begin{equation}
\phi(x)= Ce^{i\omega x}+ D e^{-i\omega x} \,,
\label{12r}
\end{equation}
and for $x\gg x_{M}$, similar solutions
\begin{equation}
\phi(x)= Fe^{i\omega x}+ Ge^{-i\omega x}\,.
\label{13r}
\end{equation}
In the expressions above, $C\,,D\,,F\,,G$ correspond to the wave amplitudes. 

At  the boundary $x=x_c \ll 0$, we impose the most general linear b.c. As is well known, this is the Robin  b.c., namely
\begin{equation}
c_1 \phi(x_c)+c_2\phi'(x_c)=0 \,,
\label{14r}
\end{equation}
$c_1\,,c_2$ being constants.
Thus, one has for $x_c <x\ll 0$,
\begin{equation}
\phi(x)= C e^{i\omega x}+ D e^{-i\omega x} \,,\quad C(\omega)= \frac{D}{ Q(\omega)} e^{-2i\omega x_c}\,,
\label{15r}
\end{equation}
where
\begin{equation}
Q(\omega)=\frac{i\omega c_2+c_1}{i\omega c_2-c_1}  \,.
\label{1qr}
\end{equation}
In analogy with the scattering theory,  it is convenient to introduce  the $M$-matrix, a $2 \times 2$ matrix which maps the vector $( C, D)$ into the vector $(F, G)$. As a consequence, the entries of the matrix M will depend on the frequencies $\omega$. 

With the Sommerfeld b.c. stipulating the absence of outgoing waves, one has $G=0$. As a result, one obtains the relation,
\begin{equation}
\frac{M_{22}}{M_{21}}=\frac{1}{Q(\omega)} e^{-2i\omega x_c} \,. 
\label{Z}
\end{equation}
The ratio $R(\omega)=M_{21}/M_{22}$ is the reflection coefficient of the barrier so Eq.~\eqref{Z} is essentially identical to that reported in \cite{Bueno}.
Once $M_{22}$ and $M_{21}$ are known, the above relation permits the computation of the frequencies $\omega$, which in general are complex numbers.

The same result can be obtained starting from  the Green function (kernel of the resolvent related to perturbation operator). The Sturm-Liouville method gives, 
\begin{equation}
G(x,y)= \frac{1}{W} \left( N(x,y)+\frac{f(x)f(y)}{M_{22}e^{2i\omega x_c}Q(\omega)-M_{21}}  \right)\,,
\label{ZZ}
\end{equation}
where
\begin{equation}
N(x,y)=  \theta(y-x)f(x)g(y)+  \theta(x-y)f(y)g(x)\,,
\label{ZZ1}
\end{equation}
with $Lf(x)=0$, satifying the conditions
\begin{equation}
f(x)=e^{i\omega x}\,,\, x \rightarrow \infty \,, \quad f(x)=
Ce^{i\omega x}+ D e^{-i\omega x} \, x \rightarrow- \infty\,,
\label{f}
\end{equation}
and  $Lg(x)=0$, satifying the conditions
\begin{equation}
g(x)=e^{-i\omega x}\,,\, x \rightarrow -\infty \,, \quad g(x)=
Fe^{i\omega x}+ G e^{-i\omega x} \, x \rightarrow \infty\,.
\label{g}
\end{equation}
Here, $W$ is the constant  Wronskian associated with $f(x)$ and $g(x)$. 
It is easy to show that $L G(x,y)=\delta(x,y)$, and the Sommerfeld B.C. at $x=\infty$, as well as  the Robin B.C. at $x_c$ are satisfied. Thus, $G(x,y)$ 
is the correct Green function of the Sturm-Liouville problem. Finally, one notes that the  $G(x,y)$ is singular when the condition (\ref{Z}) is satisfied. 
This is in agreement with Refs.~\cite{Mark,Bueno}.

When one is dealing with a BH, $x_c=x_H=-\infty$, and since $\mbox{Im}(\omega) >0 $, the QNM condition reduces to
the usual one
\begin{equation}
M_{22}= W=0 \,.
\label{ZBH}
\end{equation}
The entries of $M$-matrix may be evaluated numerically or by WKB approximation or other approximations. A complete treatment of WKB approximation can be found in Refs.~\cite{kono11, IW}. \\
The leading WKB approximation, as reported  in Quantum Mechanics books like for example \cite{Merza}, takes the form
\begin{equation}
 M_{11}=M_{22}=\theta+\frac{1}{4\theta}\,, \quad
 M_{12}=-M_{21}=i\left(\theta-\frac{1}{4\theta}\right)\,,
\label{MG}
\end{equation}
where $\theta$ is given by, for frequencies below the top of the barrier,
\begin{equation}
\theta=e^{\alpha(\omega)}\,, \quad \alpha(\omega)=\int_{x_1}^{x_2} dx\sqrt{V(x)-\omega^2} \,.
\label{t}
\end{equation}
Here, $x_1$ and $x_2$ are the classical turning points of the effective potential barrier (where $V(x_{1,2})=\omega^2$).  Above the barrier $\omega^2>V(x)$ and the radical under the integral must be taken with opposite sign, making $\theta$ apparently real. However, the turning points then move away from the real axis to take complex values, and the WKB result must be treated with greatest care. For example, in a parabolic barrier the turning points are complex conjugate, purely imaginary,  numbers.

The quantity $\alpha(\omega)$ can be rewritten as 
\begin{equation}
  \alpha(\omega)=4(V_M-\omega^2)\int_{0}^{1} dy\sqrt{1-y^2}
  \frac{\sqrt{V_M-V(x(y))}}{-V'(x(y))} \,,
\label{t120}
\end{equation}
where $V_M\equiv V(x_M)$ is the maximum of the effective potential, and   $x=x(y)$ is defined by the inversion of the expression
\begin{equation}
 y=\frac{\sqrt{V_M-V(x)}}{ \sqrt{(V_M-\omega^2)}} \,.
\label{tt}
\end{equation}
The simplest approximation is the parabolic one, namely
\begin{equation}
V(x)=V_M-\Omega^2 (x-x_{m})^2\,,\quad \Omega^2=-\frac{V''(x_m)}{2}\,.
\label{ttt}
\end{equation}
In this case,
\begin{equation}
 \theta=e^{\frac{\pi}{2\Omega}(V_M-\omega^2)} \,.
\label{t1}
\end{equation}
Making use of the above WKB result, and recalling that
$M_{22}=\theta+\frac{1}{4\theta}$
and $M_{21}=-i(\theta-\frac{1}{4\theta})$,  in the case of  WHs or ECOs, one has
\begin{equation}
\frac{4\theta^2+1}{4\theta^2-1}=-i\frac{1}{Q(\omega)} e^{-2i\omega x_c}  =-i \frac{i\omega c_2-c_1}{i\omega c_2+c_1} e^{-2i\omega x_c} \,.
\label{ZZZ}
\end{equation}
This is a transcendental equation whose solutions gives the complex values of $\omega$, and this analytical expression  is the main result of our paper. It is important to emphasize that this is only valid for frequencies away from the peak of the barrier, since the linear matching formulas are no more strictly valid there. In particular, the imaginary part of the frequencies we are looking for should be much smaller in absolute value than the real parts, so they do not correspond to the QNM observed by LIGO which have a damping time of order few \mbox{ms}, the same as the period of the ringdown frequencies. 

There are few regimes were this formula is worth exploring. The most important regime for us corresponds to  $\theta\gg 1$ and levels well below the peak of the barrier,
where the reflection coefficient is close to one. We illustrate this case for Dirichlet and Neuman boundary conditions
for which $Q(\omega)=\pm 1$. Approximating \eqref{ZZZ} we get
\begin{equation}\label{echo0}
1+\frac{1}{2\theta^2}\simeq \pm  i\, e^{-2i\rm{Re}\omega\,x_{c}}e^{2\rm{Im}\omega\,x_{c}}\,,
\end{equation}
which taking the logarithm has the solutions (since $\theta$ is real)
\begin{equation}
\mathrm{Re}(\omega_{n})=\frac{\pi(n \pm 1/4)}{x_c}, \quad
\mathrm{Im}(\omega_{n})\,=\frac{1}{4\theta_0^2\, x_c}\,.
\end{equation}
Since in our approximation $\mathrm{Im}(\omega_n)$ is very small, above we have put
\begin{equation}\label{echo}
\theta_0^2=\text{e}^{-\frac{\pi}{\Omega}(V_M -\mathrm{Re}(\omega_{n})^2)}\,.
\end{equation}
We note that we may also write
\begin{equation}\label{echo1}
\mathrm{Im}(\omega_{n})\,=\frac{1}{4\theta_0^2\, x_c}
=\frac{1}{2T }e^{-2\int_{a}^b |p_{n}(x)|dx}\,,
\end{equation}
where $p_{n}(x)=\sqrt{(V(x)-\omega_{n}^2}$, $a,b$ are the classical turning points and $T\simeq 2x_{c}$ is the classical period for motion back and forth between the wall and the peak. When the exact formula for $T$ is used
\begin{equation}
T=2\int_{x_{c}}^a\frac{dx}{p(x)}\,,
\end{equation}
this is known as the Gamow formula for the width of the resonance.
The greater is $x_{c}$, the least dumped are the corresponding modes, with a dumping time close to the period of the classical motion of a particle trapped between the wall and the peak. 
Thus these may be interpreted as long lived modes trapped between the boundary at $x_c$ and the local maximum of the potential located near the light sphere (see also Refs.~\cite{DS,pippo,cardoso,Maselli} and references quoted therein). \\
 Given a potential, Eq.~\eqref{echo1} may be used to compute the dumping time, and an example is to be found in the table below.
Clearly these modes will cross the barrier with a reduced amplitude and a characteristic delay time with respect to the usual modes making up the ringdown signal.  They are absent for the simplest black holes considered here. It has been suggested repeatedly how these echoes might give rise to detectable effects in new born gravitational astronomy, and their detection will tell us whether  one is  dealing or not  with a BH.
However, for this one has to observe the signal source over time scales longer than the coordinate time taken  by the signal to come back and forth the reflecting wall.
For the Damour-Solodukin solution (\ref{DS}) 
with small $\lambda^2$ this may well be arbitrarily long, as observed in Ref.~\cite{DS}.   
For example,  Volkel and Kokkotas in Ref. \cite{pippo} (Table II in the appendix) have calculated the QNM frequencies for the exact
same DS model with $\lambda=10^{-4}$, obtaining very small imaginary parts for small $n$ and $l$. In Table 1, we report our numerical results for the echoes.


\begin{table}[ht]
\caption{The least dumped modes of the Damour-Solodukhin solution (\ref{DS}) with $\lambda^2=10^{-4}$ for different values of $n\,,l$ in the regime $1\ll \theta^2$.
The simplest boundary conditions with  $Q(\omega)=\pm 1$, namely $c_1=0$ or $c_2=0$, are considered.}
\center
\label{tab:buoni0}
\begin{tabular}{|c|c|c|c|c|}
\hline
 $n$ & $l$ & $Q(\omega)$ & $2M\omega$ \\
\hline
1 & 1 & 1 & $ -0.426368 - 0.00377225 i$ \\
1 & 2 & 1 & $ -0.426368 - 0.0000808075 i $\\
2 & 1 & 1  &  $ -0.767462 - 0.583787 i $\\
2 & 2 & 1  & $ -0.767462 - 0.00215834 i $\\
1 & 1 & -1 & $-0.255821 - 0.000893275 i $ \\
1 & 2 & -1 &  $ -0.255821 - 0.0000316106 i $\\
2 & 1 & -1  &  $ -0.596915 - 0.0327359 i$\\
2 & 2 & -1  & $ -0.596915 - 0.000330279 i$\\
\hline
\end{tabular}
\end{table}
We see that the imaginary parts are much smaller than the real parts, in contrast with the usual QNM characterizing the ringing phase of the black holes.\\
The second regime where the resonance frequencies are well above the barrier seems to be unexplored\footnote{The formula given above is probably correct for this case too, but we have not investigated this.}, and there again there is available a WKB formula\cite{popov}. For the simple case of a potential barrier the turning points are at the complex values $\pm i(\omega^2-V_{M})/\Omega$, and we get
\[
\theta=\exp\left(\frac{i\pi}{2\Omega}(\omega^2-V_{M})\right)
\]
Thus $|\theta|\ll1$ and the expansion \eqref{echo0} is invalid. Complex solutions will be investigated elsewhere, but one thing we can say is that the transmission coefficients of these modes will be larger than for modes below the barrier.\\
Finally 
values such that $\theta$ is real and of order $\theta^2\simeq -1$, correspond to the turning points very close to each other and the WKB formula as given is invalid. In this case an improved WKB formula has been provided by Connors\cite{connor:1973} and used successfully in molecular physics. It probably requires numerical evaluation. \\
However these values are close to the  limit $x_c \rightarrow -\infty$, namely the  BH case, and it is interesting that the frequencies they give
\begin{equation}
  \omega^2=V_M-2i\Omega\left(n+\frac{1}{2}\right), \quad n=0,1,2,...,
  \mbox{Re}( \omega)<0\,,\quad n=-1,-2,..., \mbox{Re} (\omega) >0\,
\label{2t}
\end{equation}
are still compatible with the leading quadratic  approximation \cite{M,SW} (see also \cite{VZ}), but of course the real part is not a fixed constant in the correct solution.  The imaginary parts fit very well.\\
To conclude, we emphasise once more the asymptotic completeness theorem of Lax-Phillips and its generalisation by Bachelot and Bachelot: the asymptotic signal at large times is well approximated by
\begin{equation}
\Phi(t,x)\approx\sum_{j}C_{j}e^{-i\omega_{j}t}f_{j}(x,\omega_{j}), \qquad t\to+\infty
\end{equation}
where the sum includes all resonant frequencies and Jost modes, under and above the barrier, plus the anti-bound states. We computed the least dumped modes under the barrier, and indicated the need to compute the remaining part, completing half of the full problem. The other half, that of computing the Jost functions, is to be done.



\section{Hawking radiation via the tunneling method}

In this Section we will investigate the presence of Hawking radiation in the static solutions with horizon we have discussed so far. Of course for astrophysical bodies the Hawking radiation is not observationally important due to the large masses involved, but nevertheless it has great theoretical significance. 
There exist many approaches to discuss the Hawking radiation phenomenon~\cite{HT}. Here,
we follow the so called tunneling method introduced by Parik and Wilczek~\cite{parikh} in its covariant variant dubbed HT tunneling method~\cite{vanzo, Angh}. 
Within the covariant tunneling method, if one makes use of coordinates system in which the metric is regular and the vector field $\frac{\partial }{\partial r}$ is
past directed, the range of $r$ is $0 <r < \infty$. In this way, one simplifies the derivation. We will choose the Painlev\`e gauge (\ref{pg}).   
To start with, we recall that the action of a massless test particle is given by
\begin{equation}
I=\int_\gamma \partial_\mu I dx^\mu\,,\label{action}
\end{equation}
where $\gamma$ represent a path crossing the horizon. 
This implies that the most appropriate set of coordinates is the one which is regular on the horizon and with $\rho=r$ (we remind that
$r$ denotes the areal radius of the metric). The action satisfies the Hamilton-Jacobi (HJ) relativistic equation
\begin{equation}
g^{\mu\nu}\partial_\mu I\partial_\nu I=0\,, \label{var}
\end{equation}
where $g_{\mu\nu}$ is the metric tensor. Once the angular coordinates are kept fixed, the radial trajectory of a massless particle is described by HJ relativistic equation,
\begin{equation}
\gamma^{a b} \partial_aI  \partial_b I=0\,.\label{nulltrajectory}
\end{equation}
In the Painlev\`e regular gauge (\ref{pg}), and introducing the particle energy $E=-\partial_vI$, one has
\begin{equation}
\partial_r I=\frac{2E}{\sqrt{A B}}\,.
\end{equation}
In this regular gauge (see Refs.~\cite{noi,vanzo}),
\begin{equation}
I=2 E\int_\gamma \frac{1}{\sqrt{A B}}dr\,, 
\end{equation}
in which the integration variable $r$ is crossing the horizon.
Let us assume the existence of a trapping horizon located at $r=r_H$, such that from (\ref{Hcond}) we get,
\begin{equation}
B(r_H)=0\,,\quad B'_{H}>0\,.\label{horizon}
\end{equation}
We can write, near to the horizon,
\begin{equation}
B(r)=B'_H(r-r_H)+...\,.
\label{nh}
\end{equation}
 One may split the integration over $r$ in 
 three contributions: one in a neighbor of $r_H$, the other two in which $r$ is different with respect to $r_H$. 
 In the near horizon approximation, one has
\begin{equation}
I=2E\int \frac{dr}{\sqrt{A(r)}\sqrt{B'_H(r-r_H)}}+I_1\,. 
\end{equation}
If one is dealing with a BH, one finds,  
\begin{equation}
A(r)=A'_H(r-r_H)+...\,.
\label{nh1}
\end{equation}
As a consequence,
\begin{equation}
I=2E\int \frac{dr}{\sqrt{A'_HB'_H}(r-r_H-i\varepsilon)}+I_1\,. 
\end{equation}
In the expressions above, $I_1$ is a {\it real}  finite contribution and in the first integral the horizon divergence is present and it has to be cured by deforming in a 
suitable way the  integration path according to Feynman prescription for the relativistic propagator (see Refs.~\cite{noi, vanzo}). 
In this case, an imaginary part of the action is {\it present} and reads,
\begin{equation}
\mbox{Im} I=2 \pi E  \frac{1}{\sqrt{A'_HB'_H}}\,. 
\end{equation}
Within the  WKB approach, the tunneling probability  is given by
\begin{equation}
\Gamma=e^{-2 Im[ I]}\,,
\end{equation}
and one has 
\begin{equation}
\Gamma=\exp\left(- \frac{4 \pi E}{\sqrt{A'_HB'_H}}\right)\,. 
\end{equation}
Thus, the Hawking temperature will be~\cite{Angh},
\begin{equation}
T_H =\frac{\sqrt{A'_HB'_H}}{4 \pi}\,. 
\end{equation}
On the other hand, if one is dealing with a WH, there is a trapping horizon, but $A_H$ is {\it not vanishing} and positive. 
Thus, in the first integral, an integrable singularity is present, and no imaginary part appears, a strong indication that  {\it there is  no Hawking effect in static WH}. It is easy to show that this conclusion is gauge independent. Quantum effects of the type envisaged by Hawking are of course to be expected for dynamical, time dependent wormholes.

\section{Conclusions}

In this paper we have presented  an unified approach for the study of black holes, wormholes and horizonless compact objects. We have mainly 
investigated the Static Spherically Symmetric case.  We have investigated the ringdown phase of QNMs and we have derived the analytical expressions ~\eqref{ZZZ}-\eqref{echo1}, 
which are  the main result. They can be used to reconstruct the complex frequencies of the signal via the Green function method. We have discussed some regimes and emphasized
the approximate character of the WKB treatment. In the case of wormholes and horizonless compact objects,  in the first regime, we have provided an analytical expression for
the complex frequencies,  which shows the presence of echoes, here defined as long lived modes trapped between the inner wall and the potential barrier. A numerical table 
confirms the compatibility of our results with the numerical simulations presented in literature. Furthermore, a correction to the Schutz-Will formula, valid 
for  wormholes and horizonless compact objects, is also presented. Note that a similar WKB approximation for long-lived modes appeared in Ref.~\cite{ECO4}.

From the phenomenological point of view, the most interesting cases are the rotating ones. With regard to this issue, we observe that the ring down phase analysis in the rotating case may be reduced to a perturbation equation which is formally described by a Schr\"{o}dinger operator of the type discussed in Section {\bf 4}, the only difference being that the effective
potential now contains additional terms depending on the angular momentum. This has been recently considered in Ref.~\cite{ Conk, Ma,Bueno,last2,last3}). It should be
stressed that this is true only when the background spacetime in the exterior of the ultra compact object is the Kerr spacetime. In the case of a generic rotating gravastars, 
in general, the exterior metric will not be a Kerr metric, see  \cite{Gla}. A similar situation is to be expected in the rotating wormhole case. 

If we include quantum effects, making use of the tunneling method, we have shown that the Hawking radiation is absent in the static WH case. Finally, using the results of the  paper~\cite{Angh}, the absence of Hawking radiation is conjectured valid in the stationary WH case as well.

Furthermore, we have investigated  some explicit fluid models describing BHs, ECOs and WHs in the framework of GR. In the specific, by using a reconstruction technique, 
we have shown that
interesting regular BH  or horizonless solutions can be found without invoking exotic fluids with negative effective equation of state parameter, avoiding in 
this way the well known instabilities. An interesting investigation of QNMs and greybody factors in several well-known problems in 2+1 and 3+1 space-times can be also found in Ref.~\cite{Rincon}.


\subsection*{Appendix A: Scalar tensor models}

Another way to deal with possible existence of wormholes is to consider scalar tensor models. We shall restrict the discussion 
to GR plus a non self-interacting massless scalar field. For a general minimal coupled scalar field see the recent review by Bronnikov in Ref.~\cite{bronnikov}.

The first important example is the celebrated Brans-Dicke theory \cite{BD,CL,AC,FVZ}, investigated in a more general contest in  the recent paper \cite{Faraoni}. For a geometric approach see \cite{Toller}. 
The action reads
\begin{equation}
  I_{BD}=\frac{1}{16 \pi} \int \sqrt{-g} \left( \Phi R-\frac{\omega}{\Phi}\partial_\mu \Phi\partial^\mu \Phi
  \right)\label{bd}\,.
\end{equation}
This theory admits a family of SSS  solutions given by
\begin{equation}
ds^2=- W(\rho)^{\frac{1+\varepsilon}{\gamma}}dt^2+\frac{d\rho^2}{ W(\rho)^{1+\varepsilon}}+r^2(\rho)dS^2\,,
\label{bdm}
\end{equation}
where
\begin{equation}
W(\rho)=\left(1-\frac{2M}{\rho}\right)\,, \quad M=m\sqrt{(1+\gamma)/2}\,,
\label{w}
\end{equation}
and the areal radius
\begin{equation}
r(\rho)=\rho W(\rho)^{-\varepsilon}\,.
\label{w1}
\end{equation}
In the expressions above,
\begin{equation}
\gamma=\frac{\omega+1}{\omega+2}\,, \quad \varepsilon=\frac{\sqrt{2} \gamma}{\sqrt{(1+\gamma)}}-1\,.
\label{w2}
\end{equation}
The Brans-Dicke scalar field results to be
\begin{equation}
\Phi(\rho)=\Phi_0 W(\rho)^{-\frac{1}{(\omega+2)\sqrt{(1+\gamma)2}}}\,.
\label{bds}
\end{equation}
The trapping scalar reads
\begin{equation}
\chi=W(\rho)^{1+\varepsilon} (\frac{dr}{d\rho})^2=W(\rho)^{-1}\left(\rho-2M-\varepsilon \right)\,.
\label{tbd}
\end{equation}
Thus, since $\rho > 2M$, if $\varepsilon >0$, then there exists a trapping horizon located at
\begin{equation}
\rho_H= 2M+\varepsilon\,.
\label{tbd1}
\end{equation}
Since $A_H=W(\rho_H)^{\frac{1+\varepsilon}{\gamma}}>0  $, one is dealing with a WH~\cite{AC,FVZ}.
Now, $\varepsilon > 0$ is equivalent to $\gamma >1$, namely $ \omega <0$. 
This means that in order to deal with a WH, the Brans-Dicke field has to be a ghost, namely with  the wrong sign in the kinetic term.

If $\omega >0$, then $\gamma <1$, and there exists a naked singularity, 
since there is no horizon. 
Note that in the limit $|\omega|\rightarrow \infty$, $\gamma\rightarrow 1$ and $\varepsilon\rightarrow 0$ and we recover the GR BH solution.

The minimally coupled case has been investigated and rediscovered in several papers, 
and here we report only the case of a massless scalar ghost field (wrong sign of the canonical kinetic term) in the parametrization found by Bronnikov~\cite{bronnikov}.
The static solution is of general form with
\begin{equation}
r=k^2 \frac{k e^{m\rho}}{\sin k\rho}\,,
\label{e1}
\end{equation}
and
\begin{equation}
B=A=\frac{\sin^2 k\rho}{r^2 k^2}\,.
\label{w11}
\end{equation}
Here, $m$ and $k$ are positive real parameters.
It is easy to show that
\begin{equation}
\chi=\frac{k}{k^2 r^2\sin^2 k \rho }\left(m \sin k\rho - k \cos k \rho \right)\,.
\label{tbd1fine}
\end{equation}
Thus, the trapping double horizon is located in the correspondence of $\tanh k\rho_H=\frac{k}{m}$.

\subsection*{Appendix B: Photosphere}

It is interesting to discuss the properties of null geodesics in the class of static SSS with metric (\ref{sgen0}). 
To begin with, we first derive the photosphere equation.
We can set $\theta=\frac{\pi}{2}$ and choose as affine parameter describing a massless particle the angle $\phi$. 
Thus, the Lagrangian of a massless test particle reads
\begin{equation}
L=\frac{1}{2V}\left(-A(\rho)\dot{t}^2+\frac{\dot{\rho}^2}{B(\rho)}+ r^2(\rho)\right)\,,
\label{sgen000}
\end{equation}
where $\dot{t}=\frac{dt}{d\phi} $, $\dot{\rho}=\frac{d \rho}{d\phi} $ and $V$ is the einbein field which implements the reparametrization invariance of the 
massless particle action. The equation of motion associated to $V$ leads to
\begin{equation}
A(\rho)\dot{t}^2=\frac{\dot{\rho}^2}{B(\rho)}+ r^2(\rho)\,.
\label{1}
\end{equation}
Then there are two first integrals, the first due to the fact that the Lagrangian is time independent, namely
\begin{equation}
A(\rho)\frac{(\dot{t})^2}{V}=E\,,
\label{2}
\end{equation}
the other one related to the conservation of the angular momentum
\begin{equation}
h=\frac{1}{2V}\left(-A(\rho)\dot{t}^2+\frac{\dot{\rho}^2}{B(\rho)}- r^2(\rho) \right)\,.
\label{3}
\end{equation}
As a result,
\begin{equation}
\left(\frac{d \rho}{dr}\right)^2 \dot{u}^2+B u^2=J^2\frac{B}{A}  \,,
\label{4}
\end{equation}
where $u=1/r$, and $J^2=\frac{E^2}{h^2}$. For a fixed radius of the photosphere, $u=u_0$,
$J^2$ is 
\begin{equation}
A_0 u_0^2=J^2  \,.
\label{5}
\end{equation}
The equation of motion with respect to $u$ is derived as
\begin{equation}
  2\left(\frac{d \rho}{dr}\right)^2\ddot{u}+ \dot{u}\frac{d}{d \phi}\left(\frac{d \rho}{dr}\right)^2
  +2B u+u^2 \frac{d B}{d u}=J^2 \frac{d(\frac{B}{A})}{du}  \,.
\label{6}
\end{equation}
As a result, from the above equation and (\ref{5}) with constant $u_0$,  one obtains
\begin{equation}
  2A_0 +u_0 \frac{d A_0}{d u}=0  \,.
\label{7}
\end{equation}
For example, for the Schwarzschild BH,  $A(u)=(1-ur_H)$. Thus, $u_0=\frac{2}{3 r_H}$, a well known result.

In the general static gauge, one has
\begin{equation}
  2A_0 \left(\frac{d r}{d \rho}\right)_0=r_0 \left(\frac{d A}{d \rho}\right)_0  \,.
\label{8}
\end{equation}

\subsection*{Appendix C: The tortoise in BH and WH cases}

Now we would like to discuss the difference between a BH and a WH in the Ellis gauge. The key point, as we have already stressed, is  
the different nature of the trapping horizon. We may start making use of the proper distance (Ellis) gauge, with the proper distance defined by
\begin{equation}
\sigma=\int \frac{d\rho}{\sqrt{B}}+\sigma_0\,.
\label{et}
\end{equation}
We can always choose the constant $\sigma_0$ such that the horizon is located at $\sigma_H=0$. 
In this form, the coordinate $\sigma$ may be extended to the whole real line, due to the fact that the metric remains regular.

As a specific but important example, let us consider the variant of the Damour-Solodukin WH~\cite{DS}, 
namely a static SSS with $\rho=r$, $A=1-\frac{2m_1}{r} $, and $B=1-\frac{2m}{r}$, with $m_1=m(1-b^2)$, $m\,,b$ (positive) constants. Thus, $m_1 <m$ and
\begin{equation}
ds^2=-\left(1-\frac{2m_1}{r}\right)dt^2+\frac{dr^2}{\left(1-\frac{2m}{r}\right)}+ r^2 dS^2\,.
\label{sgends}
\end{equation}
Since $m> m_1$ , at $r_H=2m$, we recover  $A_H=\frac{m-m_1}{m}=b^2$, and
\begin{equation}
\sigma=\sqrt{r^2-2 m r}+M\ln \left(\frac{r-m+\sqrt{r^2- 2r m}}{m}   \right)\,.
\label{t2}
\end{equation}
Note that the horizon corresponds to $\sigma_H=0$.

Now let us consider the near horizon approximation with $r=2m+\varepsilon$. It follows that $\sigma^2=8m\varepsilon$, and the near horizon metric  can be recast in the form
\begin{equation}
ds^2=-\left(a^2_0+a^2_1 \sigma^2 \right)dt^2+d\sigma^2+\left(r^2_0+r_1^2 \sigma^2\right) dS^2\,.
\label{t5}
\end{equation}
If $a_0$ is not vanishing, one has a WH in the near horizon approximation, otherwise we deal with a BH. As a result, one can see that the near horizon metric is
regular also for $\sigma<0$. 

The  tortoise coordinate is given by
\begin{equation}
  \rho^*=\int \frac{d\sigma}{\sqrt{ a^2_0+a^2_1\sigma^2 }}=\frac{1}{a_1}\ln\left(a_1\sigma+
  \sqrt{ a^2_0+a^2_1\sigma^2 } \right)\,.
\label{t7}
\end{equation}
Thus, it follows that $\rho^*$ assumes also negative real values, and on the horizon where $\sigma_H=0$ one has
\begin{equation}
\rho_H^*=\frac{1}{a_1}\ln \left( \sqrt{ a^2_0 } \right)\,.
\label{t8}
\end{equation}
In the case of a WH, this is a finite value. If one is dealing with a BH with $a_0=0$, one gets the  well known result $\rho_H^*\rightarrow-\infty$.

Within the near horizon approximation, the effective potential in (\ref{effp}) reads
\begin{equation}
V(\sigma)=V_0+V_1 \sigma^2\,,
\label{t9}
\end{equation}
where
\begin{equation}
V_0= \left( \frac{l(l+1)}{r_0^2} +\frac{r_1^2}{r_0^2}\right) a_0^2\,,
\label{t10}
\end{equation}
and
\begin{equation}
V_1=\left(\frac{l(l+1)}{r_0^2}+\frac{2r_1^2}{r_0^2}\right) \left( a^2_1-\frac{a_0^2r_1^2}{r_0^2}\right)\,.
\label{t100}
\end{equation}
Thus, in the case of WH with $a_0^2\neq 0$, $V(\sigma)$ has {\it a local minimum} at $\sigma_H=0$, as soon as $V_1>0$, namely when
$ a^2_1>\frac{a_0^2r_1^2}{r_0^2} $. The local minimum is present also in the expression $V(\rho^*$), since $\rho^*$ is a monotonic function with respect to $\sigma$.

Let us investigate other stationary points of effective potential. In the case of BH, it is well known that there exists a local maximum. The simplest way to see it
is to note that, for large values of $l$, the effective potential reduces to
\begin{equation}
V(\rho)=\frac{l(l+1) A(\rho)}{r^2(\rho)}\,.
\label{rwesc00}
\end{equation}
This leading term has a local maximum when
\begin{equation}
  2A_0 \frac{dr}{d \rho}_0= r_0 \left(\frac{d A}{d \rho}\right)_0  \,.
\label{7qn}
\end{equation}
An important remark is that this is the same condition which defines the unstable photon sphere orbit at $r=r_0$.
In the case of BH, this local maximum, located approximately around the photon sphere, is the only stationary 
point, because at the trapping horizon with $\rho_H^*=-\infty$ the effective potential is vanishing.

On the other hand, the situation in the case of WH is completely different, since we have shown that at the trapping horizon, or mouth, there is a local minimum at
$\rho^*_H=0$. Since the effective potential is still vanishing at $\rho^*\rightarrow-\infty$, it follows that there exists another symmetric local maximum, located at 
$\rho^*=-\rho_0^*$~\cite{DS}.

\end{document}